\documentclass[11pt,dvips]{article}
\usepackage{epsfig}
\newenvironment{descit}[1]{\begin{quote} \textit{#1}}{\end{quote}}

\begin{document}

\title{Evaluating Recommendation Algorithms\\
by Graph Analysis}
\author{Batul J. Mirza, Benjamin J. Keller, and Naren Ramakrishnan\\
Department of Computer Science\\
Virginia Tech, VA 24061\\
Email: \{bmirza,keller,naren\}@cs.vt.edu}
\date{}
\maketitle

\begin{abstract}
\noindent
We present a novel framework for evaluating recommendation algorithms
in terms of the `jumps' that they make to connect people to artifacts.
This approach emphasizes reachability via an algorithm within the implicit
graph structure underlying a recommender dataset, and serves as a complement
to evaluation in terms of predictive accuracy. The framework allows us
to consider questions relating algorithmic parameters to properties of
the datasets. For instance, given a particular algorithm `jump,' what
is the average path length from a person to an artifact? Or, what choices
of minimum ratings and jumps maintain a connected graph? We illustrate the
approach with a common jump called the `hammock' using movie
recommender datasets.\\


\end{abstract}
\thispagestyle{empty}

\newpage

\setcounter{page}{1}


\section{Introduction}
\label{intro}

Recommender systems \cite{Resnick1} constitute one of the fastest
growing segments of the Internet economy today. They help reduce
information overload and provide customized information access for
targeted domains. Building and
deploying recommender systems has matured into a
fertile business activity, with benefits in retaining customers and
enhancing revenues. Elements of the recommender landscape include
customized search engines, handcrafted content indices, personalized shopping
agents on e-commerce sites, and news-on-demand services.
The scope of such personalization thus extends to many
different forms of information content and delivery, not just web
pages. The underlying algorithms and techniques, in turn, range from
simple keyword matching of consumer profiles, collaborative filtering,
to more sophisticated forms of data mining, such as clustering web
server logs.

Recommendation is often viewed
as a system involving two modes (typically people and artifacts,
such as movies and books) and has been studied in domains that
focus on harnessing online information resources, information
aggregation, social schemes for decision making, and user
interfaces. A recurring theme among many of these applications 
is that recommendation is implicitly cast as a task 
of learning mappings (from people to recommended artifacts, for example) or of 
filling in entries to missing cells in a matrix (of consumer preferences, for 
example). Consequently, recommendation algorithms are evaluated by
the accuracies of their predicted ratings. We approach recommendation from 
the different and complementary perspective of considering the
connections that are made.

\subsection{Motivating Scenarios}
We describe three scenarios involving recommender system design to motivate
the ideas presented in this paper.

\begin{itemize}
\item {\it Scenario 1:} A small town bookstore is designing 
a recommender system to provide targeted personalization for its customers. 
Transactional data, 
from purchases cataloged over three years, is available. The store
is not interested in providing specific recommendations of books, but
is keen on using its system
as a means of introducing customers to one another and 
encouraging them to form reading groups.
It would like to bring sufficiently concerted groups of people together, based
on commonality of interests.
Too many people in a group would imply a diffusion of interests; modeling
reading habits too narrowly might imply that some people cannot be matched 
with anybody. How can the store relate commonality of interests to the
sizes of clusters of people that are brought together?
\item {\it Scenario 2:} An e-commerce site, specializing in books, CDs, 
and movie videos, is installing a recommendation service.
The designers are acutely aware that people buy 
and rate their different categories of
products in qualitatively different ways.
For example, movie ratings follow a {\it hits-buffs}
distribution: some people (the buffs) see/rate almost all movies, and some
movies (the hits) are seen/rated by almost all people. Music CD ratings are
known to be more clustered, with hits-buffs distributions visible only within
specific genres (like `western classical'). Connections between different
genres are often weak, compared to connections within a genre.
How can the designers reason about and visualize the structure of 
these diverse recommendation spaces, to allow them
to custom-build their recommendation algorithms?

\item {\it Scenario 3:} An online financial firm is investing in a 
recommendation service and is requiring
each of its members to rate at least $\kappa$ products of
their own choice to
ensure that there are enough overlaps among ratings. The company's research
indicates that people's ratings typically follow
power-law distributions. Furthermore, the company's marketers have decided
that recommendations of ratings can be `explainably
transferred' from one person to
another if they have at least $6$ ratings in common. Given these statistics
and design constraints, what value of $\kappa$ should be set by the company
to ensure that every person (and every product)
is reachable by its recommendation service? 

\end{itemize}

\noindent
The common theme among these applications is that they emphasize many
important aspects of a recommender system, other than predictive accuracy:
its role as an indirect way of bringing people together, 
its signature pattern of making connections, and
the explainability of its recommendations. To
address the questions raised by considering these aspects of recommendation,
we propose a framework based on a mathematical model of the social
network implicit in recommendation. This framework allows a more
direct approach to evaluating and reasoning about recommendation
algorithms and their relationship to the recommendation patterns
of users. We effectively ignore the issue of predictive accuracy, and so
the framework is a complement to approaches based on field studies.

%

\subsection{Reader's Guide}
Section~\ref{back}
surveys current research and
motivates the need for a new approach to analyzing algorithms for
recommender systems. Section~\ref{jcmodel} introduces the `jumping
connections' framework and develops a mathematical model based on random
graph theory. Section~\ref{results} provides experimental results for one
particular way of `jumping' on two application datasets. Issues related to
interpretation of results from our model are also presented here. Finally,
Section~\ref{future} identifies some opportunities for future research.

\section{Evaluating Recommendation Algorithms}
\label{back}
Most current research efforts cast recommendation
as a specialized task of information retrieval/\hskip0ex filtering or
as a task of function approximation/\hskip0ex learning mappings 
\cite{aggarwal1,
basu1,billsus,
eigentaste,good,herlocker,
hill,kitts,konstan1,
pennock,sarwar1,sarwar3,
schafer1,
ringo1,soboroff,
terveen}. Even approaches
that focus on clustering view clustering primarily as a
pre-processing step for functional modeling \cite{kohrs1}, or as a 
technique to ensure scalability \cite{Conner1,sarwar3}
or to overcome sparsity of ratings \cite{ungar1}. This emphasis on functional 
modeling and retrieval has influenced evaluation criteria for recommender 
systems.

Traditional information retrieval evaluation 
metrics such as precision and recall have been applied toward recommender
systems involving content-based design.
Ideas such as cross-validation on an unseen test set have been used to
evaluate mappings from people to artifacts, especially in
collaborative filtering recommender systems. Such approaches miss
many desirable aspects of the recommendation process, namely:
\begin{itemize}
\item {\bf Recommendation is an indirect way
of bringing people together.} Social network theory \cite{wasserman-faust}
helps model a recommendation system of people versus artifacts as an {\it
affiliation network} and distinguishes between a {\it primary mode} 
(e.g., people) and a {\it secondary mode} (e.g., movies), where a 
{\it mode} refers to a distinct set of entities that have similar
attributes \cite{wasserman-faust}. The purpose of
the secondary mode is viewed as serving to bring entities of the primary
mode together (i.e., it isn't treated as a {\it first-class} mode).
\item {\bf Recommendation, as a process,
should emphasize modeling connections from people to
artifacts, besides predicting ratings for artifacts.}
In many situations, users would like to request recommendations
purely based on local and global constraints on the nature of
the specific connections explored. 
Functional modeling techniques are
inadequate because they embed the task of learning a mapping from people
to predicted values of artifacts in a general-purpose learning system such
as neural networks or Bayesian classification \cite{breese}.
A notable exception is the work by Hofmann and Puzicha \cite{Hofmann}
which allows the incorporation of constraints in the form
of aspect models involving a latent variable. 
\item {\bf Recommendations should be explainable and believable.} The 
explanations should be made in terms and constructs that are natural to the 
user/application domain. It is nearly
impossible to convince the user of the quality of a recommendation obtained
by black-box techniques such as neural networks. Furthermore, it is well
recognized that ``users are more satisfied with a system that produces
[bad recommendations] for reasons that seem to make sense to them, than they
are with a system that produces [bad recommendations] for semmingly stupid 
reasons'' \cite{riloff}.
\item {\bf Recommendations are not delivered in isolation, but in the
context of an implicit/explicit social network.}
In a recommender system, the rating patterns of
people on artifacts induce an implicit social network and influence the
connectivities in this network.  Little study has been done to understand
how such rating patterns influence recommendations and how they
can be advantageously exploited.
\end{itemize}

Our approach in this paper is to evaluate recommendation algorithms using
ideas from graph analysis. In the next section,
we will show how our viewpoint addresses each of 
the above aspects, by providing novel metrics. The basic idea is to begin
with data that can be modeled as a network and attempt to infer useful
knowledge from the nodes and links of the graph. Nodes represent entities
in the domain (e.g., people, movies), and edges represent the relationships
between entities (e.g., the act of a person viewing a particular movie).

\subsection{Related Research}
The idea of graph analysis as a basis to study information networks has a long
tradition; one of the earliest pertinent studies is Schwartz and
Wood \cite{graph-schwartz}. The authors describe the use of graph-theoretic
notions such as cliques, connected components, cores, clustering,
average path distances, and the inducement of secondary graphs. The focus
of the study was to model shared interests among a web of people,
using email messages as connections. Such link
analysis has been used to extract information in many areas such as in web
search engines \cite {klein2}, in exploration of associations among
criminals \cite{lee1}, and in the field of medicine \cite {swanson1}.
With the emergence of the web as a large scale graph, interest in information 
networks has recently exploded \cite{adamic1,google,bowtie,
clever-journal,flake-kdd,
kautz1,klein2,klein1,trawling,payton1,silk,watts1}.

Most graph-based algorithms for information networks can be studied in terms
of (i) the modeling of the
graph (e.g., what are the modes?, how do they
relate to the information domain?), and (ii) the structures/operations
that are mined/conducted on the graph.
One of the most celebrated examples of graph analysis arises in search
engines that exploit link information, in addition to textual content.
The Google search engine uses the web's link structure,
in addition to the anchor text as a factor in ranking pages, based on
the pages that (hyper)link to the given page \cite{google}. Google essentially models
a one-mode directed graph (of web pages) and uses measures involving 
principal components to ascertain `page ranks.'
Jon Kleinberg's HITS (Hyperlink-Induced Topic Search) algorithm
goes a step further by viewing the one-mode web graph as
actually comprising two modes (called
hubs and authorities) \cite{klein2}. A hub is a node primarily with edges
to authorities, and so a good hub has links to many authorities. A good
authority is a page that is linked to by many hubs.
Starting with a specific
search query, HITS performs a text-based search to seed an initial set of
results. An iterative relaxation algorithm then assings hub and authority
weights using a matrix power iteration. Empirical results show that
remarkably authoritative results are obtained for search queries. The CLEVER
search engine is built primarily on top of the basic HITS
algorithm \cite{clever-journal}. The offline query-independent
computation in Google, as opposed to 
the topic-induced search of CLEVER, is one of the main reasons for the
commercial success of the former. 

The use of link analysis in recommender systems was highlighted by the
``referral chaining'' technique of the 
ReferralWeb project \cite{kautz1}. The idea is to
use the co-occurrence of names in any
of the documents available on the web to detect the existence of direct
relationships between people and thus indirectly form social networks. The
underlying assumption is that people with similar interests swarm in the
same circles to discover collaborators \cite{payton1}.

The exploration of link analysis in social structures has led to several
new avenues of research, most notably small-world networks. Small-world
networks are highly clustered but relatively sparse networks with
small average length. An example is the 
folklore notion of six degrees of separation separating any two people in
our universe: the phenomenon where a person can discover a link to any
other random person through a chain of at most six acquaintances. A small-world
network is sufficiently clustered so that most second neighbors of a node
$X$ are also neighbors of $X$ (a typical ratio would be $80\%$). On the
other hand, the average distance between any two nodes in the graph is
comparable to the low characteristic path length of a random graph. Until
recently, a mathematical characterization of such small-world networks has
proven elusive.  Watts and Strogatz \cite{watts1} provide the first
such characterization of small-world networks in the form
of a graph generation model.

\begin{figure}
\centering
\begin{tabular}{cc}
& \mbox{\psfig{figure=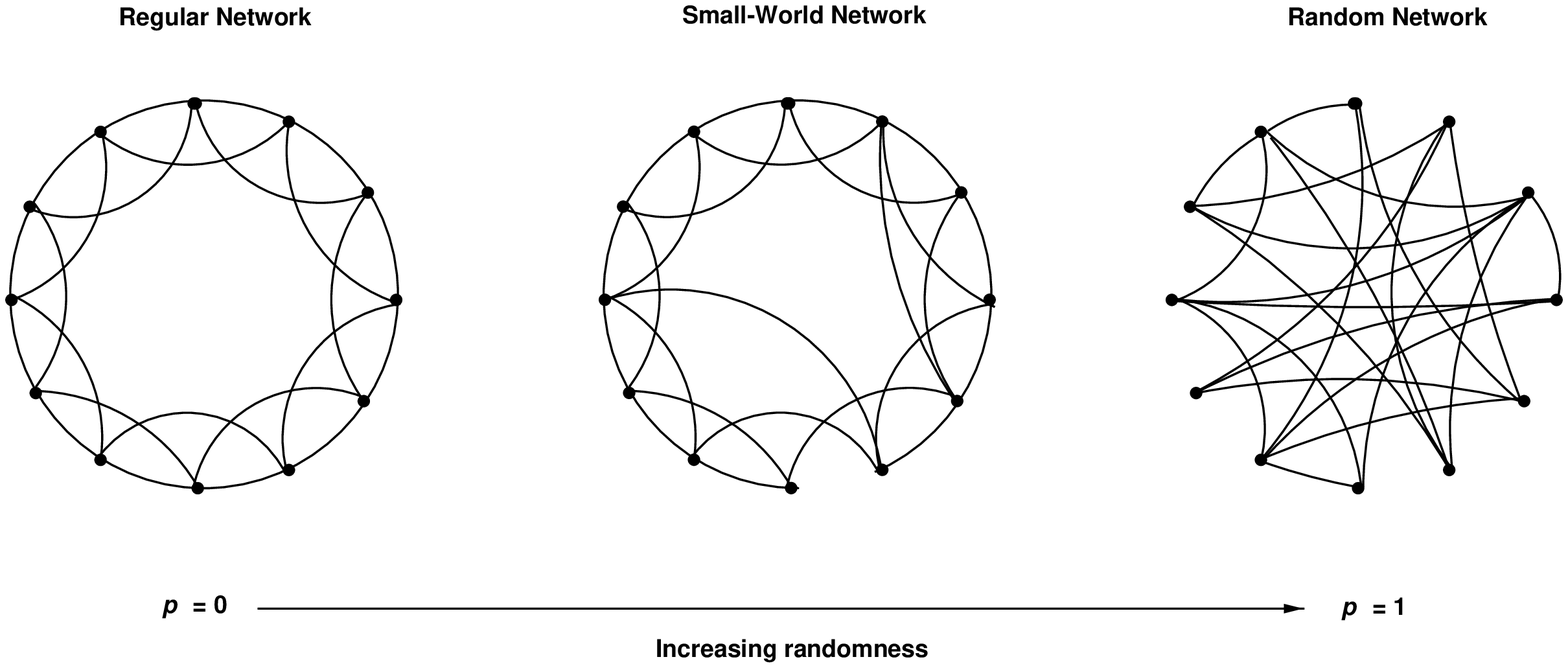,width=5in}}
\end{tabular}
\caption{Generation of a small-world network by random rewiring from a
regular wreath network. Figure adapted from \cite{watts1}.}
\label{small}
\end{figure}

\begin{figure} \centering
\begin{tabular}{cc} &
\mbox{\psfig{figure=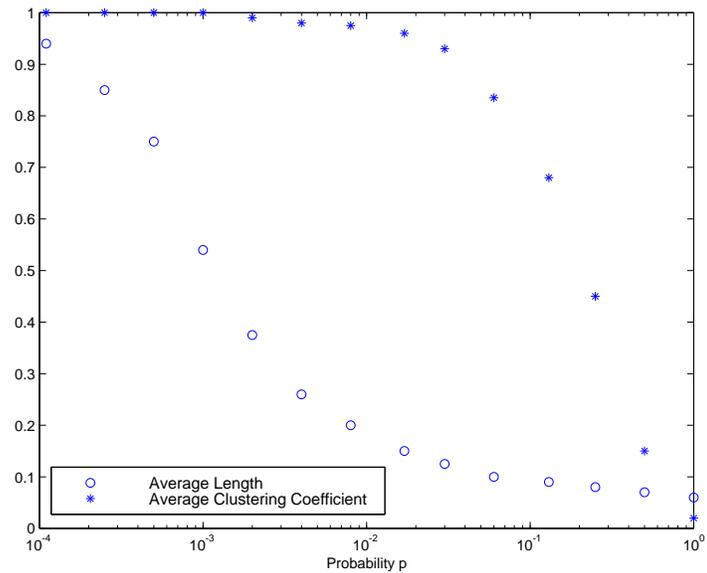,height=3in}}
\end{tabular}
\caption{Average path length and clustering coefficient versus
the rewiring probability $p$ (from \cite{watts1}). All measurements are
scaled w.r.t. the values at $p = 0$.}
\label{smgs}
\end{figure}

In this model, Watts and Strogatz
use a regular wreath network with $n$ nodes, and $k$ edges per node
(to its nearest neighbors) as a starting point for the design. A small
fraction of the edges are then randomly rewired to arbitrary points on
the network. A full rewiring (probability $p=1$) leads to a completely
random graph, while $p=0$ corresponds to the (original) wreath 
(Fig.~\ref{small}). The starting point in the figure is a regular wreath
topology of $12$ nodes with every node connected to its four nearest
neighbors. This structure has a high characteristic path length and
high clustering coefficient. The average length is the mean of the shortest
path lengths over all pairs of nodes. The clustering coefficient is
determined by first computing the local neighborhood of every node. The
number of edges in this neighborhood as a fraction of the total possible
number of edges denotes the extent of the neighborhood being a clique.
This factor
is averaged over all nodes to determine the clustering coefficient. The
other extreme in Fig.~\ref{small}
is a random network with a low characteristic path length and
almost no clustering. The small-world network, an interpolation between the
two, has the low characteristic path length (of a random network), and
retains the high clustering coefficient (of the wreath).  Measuring
properties such as average length and clustering coefficient in the region
$0 \leq p \leq 1$ produces surprising results (see Fig.~\ref{smgs}).

As shown in Fig.~\ref{smgs}, only a very small fraction of edges need to be
rewired to bring the length down to random graph limits, and yet the
clustering coefficient is high. On closer inspection, it is easy to see why
this should be true. Even for small values of $p$ (e.g., $0.1$), the result
of introducing edges between distantly separated nodes reduces not only the
distance between these nodes but also the distances between the neighbors of
those nodes, and so on (these reduced paths between distant nodes are
called {\it shortcuts}). The introduction of these edges
further leads to a rapid decrease in the average length of the network, but
the clustering coefficient remains almost unchanged. Thus, small-world
networks fall in between regular and random networks, having the small
average lengths of random networks but high clustering coefficients akin to
regular networks.

While the Watts-Strogatz model describes how small-world networks can be
formed, it does not explain how people are adept at actually finding short
paths through such networks in a decentralized fashion. Kleinberg 
addresses precisely this issue and proves that this is not possible
in the family of one-dimensional Watts-Strogatz networks 
\cite{klein1}. Embedding the notion of 
random rewiring in a
two-dimensional lattice leads to one unique model for which such
decentralization is effective.

The small-world network concept has implications for a variety of domains.
Watts and Strogatz simulate the `wildfire' like spread of an
infectious disease in a small-world network \cite{watts1}. 
Adamic shows
that the world wide web is a small-world network and suggests that
search engines capable of exploiting this fact can be more
effective in  hyperlink modeling, crawling, and finding authoritative 
sources \cite{adamic1}.

Besides the Watts-Strogatz model, a variety of models from graph
theory are available and can be used to analyze information networks.
Kumar et al.~\cite{trawling} highlight the use of traditional
random graph models
to confirm the existence of properties such as cores and connected
components in the web. In particular, they characterize the distributions 
of web page degrees 
and show that they are well approximated by power laws.
Finally, they perform a study similar to Schwartz and Wood
\cite{graph-schwartz} to find cybercommunities on the web. 
Flake et al.~\cite{flake-kdd} provide a max-flow, min-cut algorithm to 
identify cybercommunities. They also provide a focused crawling strategy to
approximate such communities.
Broder et al.~\cite{bowtie} perform a more detailed mapping of the web
and demonstrate that it has a bow-tie structure, which consists of
a strongly connected component, as well as nodes that 
link into but are not linked
from the strongly connected component, and nodes that are linked from but
do not link to the strongly connected component.
Pirolli et al.~\cite{silk} use ideas from spreading activation
theory to subsume link analysis, content-based modeling, and usage
patterns.

A final thread of research, while not centered on information networks,
emphasizes the modeling of problems and applications in ways that make them
amenable to graph-based analyses. A good example in this category is
the approach of Gibson et al.~\cite{gibson1} for mining categorical datasets.

While many of these ideas, especially link analysis,
have found their way into recommender systems,
they have been primarily viewed as mechanisms to mine or model 
structures. In this paper, we show how ideas from graph analysis
can actually serve to provide novel evaluation criteria for recommender 
systems.

\section{Graph Analysis}
\label{jcmodel}

To address the four aspects identified in the previous section, we develop a 
novel way to characterize algorithms for recommender systems.  
Algorithms are distinguished, not by the predicted ratings of
services/artifacts they produce, but by the combinations of people and
artifacts that they bring together. Two algorithms are considered
equivalent if they bring together identical sets of nodes regardless
of whether they work in qualitatively different ways.
Our emphasis is on the role of a recommender system as a
mechanism for bridging entities in a social network. 
We refer to this approach of studying recommendation as {\it jumping
connections}.

Notice that the framework does not emphasize how the recommendation is 
actually made, or the information that an algorithm uses to make connections
(e.g., does it rely on 
others' ratings, on content-based features, or both?). {\it In addition, we make
no claims about the recommendations being `better' or that they will be
`better received.'} Our metrics, hence, will not lead a designer to 
directly conclude that an algorithm $A$ is more accurate than
an algorithm $B$; such conclusions can only be made through
a field evaluation (involving feedback and reactions from users) or
via survey/interview procedures. By restricting its scope to exclude the 
actual aspect of making ratings and predictions, the jumping connections 
framework provides a systematic and rigorous way to study recommender systems.

Of course, the choice of how to jump connections will be driven by the
(often conflicting) desire to reach almost every node in the graph (i.e.,
recommend every product for somebody, or
recommend some product for everybody)
and the strength of the jumps enjoyed when two nodes are brought together.
The conflict between these goals can be explicitly expressed in our
framework.

It should be emphasized that our model
doesn't imply that algorithms only exploit
local structure of the recommendation dataset. Any mechanism --- 
local or global --- could be used to jump connections. In fact, it is
not even necessary that algorithms employ graph-theoretic notions
to make connections. Our framework
only requires a boolean test to see if two nodes are brought together.

Notice also that when an algorithm brings together person $X$ and artifact
$Y$, it could imply either a positive recommendation or a negative one. Such
differences are, again, not captured by our framework unless the mechanism
for making connections restricts its jumps, for instance, to only those
artifacts for which ratings satisfy some threshold. In other words,
thresholds for making recommendations could be abstracted into the mechanism
for jumping.

Jumping connections satisfies all the aspects outlined in the previous
section. It involves a social-network model, and thus, emphasizes connections
rather than prediction. The nature of connections jumped also aids in 
explaining the recommendations.  
The graph-theoretic nature of jumping connections allows 
the use of mathematical models (such as random graphs) to analyze the
properties of the social networks in which recommender algorithms operate.

\subsection{The Jumping Connections Construction}


We now develop the framework of jumping connections. We use concepts from
a movie recommender system to provide the intuition; this does not
restrict the range of applicability of jumping connections
and is introduced here only for ease of presentation. 

A \emph{recommender dataset} $\mathcal{R}$ consists of the ratings by a
group of people of movies in a collection.
The ratings could in fact be viewings, preferences, or other
constraints on movie recommendations.
Such a dataset can be represented as a bipartite graph 
$G=(P\cup M,E)$ where $P$ is the set of people, $M$ is the set of
movies, and the edges in $E$ represent the ratings of movies.
We denote the number of people by $N_P=|P|$, and the number of movies
as $N_M=|M|$.

We can view the set $M$ as a secondary mode that helps make
connections --- or jumps --- between members of $P$. 
A \emph{jump} is a function 
$\mathcal{J}: \mathcal{R} \mapsto S, S \subseteq P \times P$ that
takes as input a recommender dataset $\mathcal{R}$ and returns a set of
(unordered) pairs of elements of $P$. 
Intuitively, this means that the two nodes described in a given pair
can be reached from one another by a single jump.  
Notice that this definition does not prescribe how the mapping should
be performed, or whether it should use all the information present in
$\mathcal{R}$.  
We also make the assumption that jumps can be composed in
the following sense: if node $B$ can be reached from $A$ in one jump,
and $C$ can be reached from $B$ in one jump, then $C$ is reachable
from $A$ in two jumps. 
The simplest jump is the \emph{skip}, which connects two members in
$P$ if they have at least one movie in common.  

\begin{figure}
\centering
\begin{tabular}{cc}
& \mbox{\psfig{figure=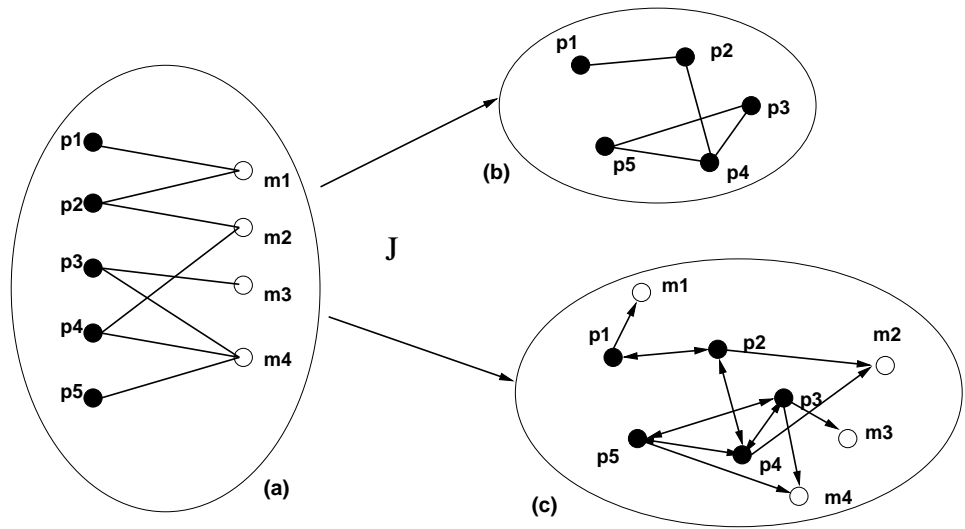,width=5in}}
\end{tabular}
\caption{Illustration of the \emph{skip} jump. (a) bipartite graph
of people and movies. (b) Social network graph, and (c)
recommender graph.}
\label{jc-intro}
\end{figure}

A jump induces a graph called a social network graph. 
The \emph{social network graph} of a recommender dataset
$\mathcal{R}$ induced by a given jump $\mathcal{J}$ is a unipartite
undirected graph $G_s=(P,E_s)$, where the edges are given by 
$E_s = \mathcal{J} (\mathcal{R})$. 
Notice that the induced graph could be disconnected based on the
strictness of the jump function. 
Figure~\ref{jc-intro} (b) shows the social network graph induced
from the example in Figure~\ref{jc-intro} (a) using a skip jump.


We view a recommender system as exploiting the social connections (the
jumps) that bring together a person with other people who have rated
an artifact of (potential) interest.
To model this, we view the unipartite social network of people as a
directed graph and reattach movies (seen by each person) such that
every movie is a sink (reinforcing its role as a secondary 
mode). 
The shortest paths
from a person to a movie in
this graph can then be used to
provide the basis for recommendations.  
We refer to a graph induced in this fashion as a \emph{recommender
graph} (Figure~\ref{jc-intro} (c)). 
Since the outdegree of every movie node is fixed at zero, paths through the
graph are from people to movies (through more people, if necessary).


The recommender graph of a recommender dataset $\mathcal{R}$ induced
by a given jump function $\mathcal{J}$ is a directed graph 
$G_r=(P \cup M,E_{sd} \cup E_{md})$, where $E_{sd}$ is an ordered set of pairs,
listing every pair from $\mathcal{J(R)}$ in both directions, and
$E_{md}$ is an ordered set of pairs, listing every pair from $E$ in
the direction pointing to the movie mode. 

\begin{figure}
\centering
\begin{tabular}{cc}
& \mbox{\psfig{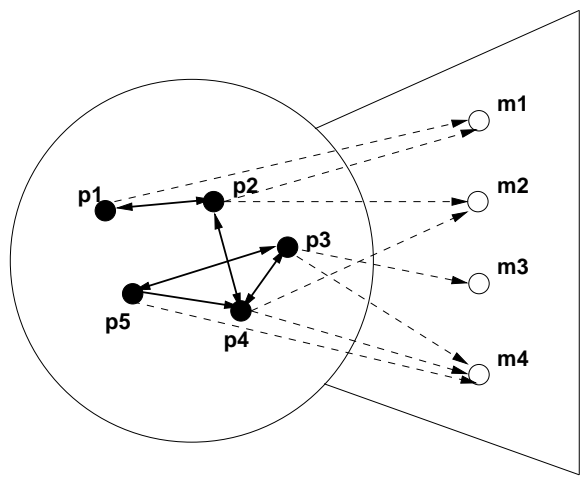}}
\end{tabular}
\caption{The jumping connections construction produces a half bow-tie graph $G_r$.}
\label{half-bowtie}
\end{figure}

\begin{figure}
\centering
\begin{tabular}{cc}
& \mbox{\psfig{figure=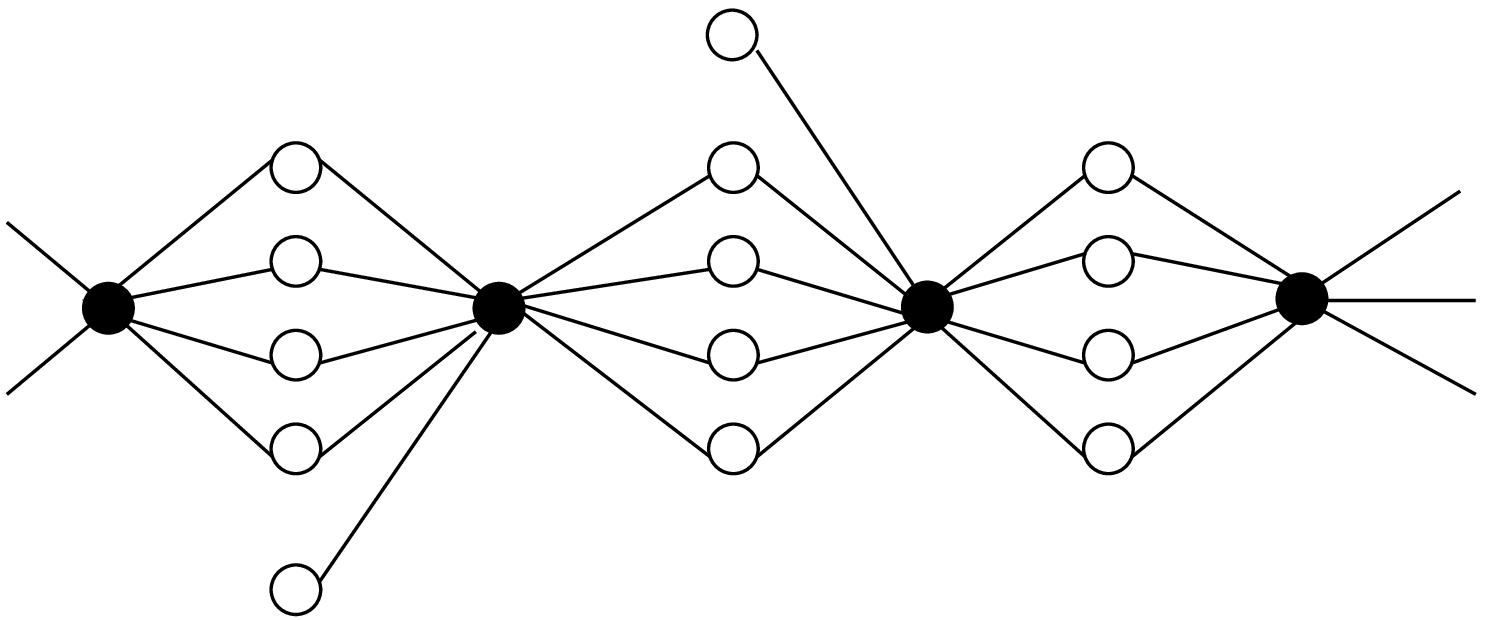,width=5in}}
\end{tabular}
\caption{A path of hammock jumps, with a hammock width
$w=4$.}
\label{hammock-pic}
\end{figure}

Assuming that the jump construction does not cause $G_r$ to be disconnected,
the portion of $G_r$ containing only people is its strongest component:
every person is connected to every other person. The movies constitute
vertices which can be reached from the strongest component, but from which
it is not possible to reach the strongest component (or any other node, for
that matter). Thus, $G_r$ can be viewed as a `half bow-tie,' 
(Figure~\ref{half-bowtie}) as contrasted to the full bow-tie nature of the web,
observed by Broder et al.~\cite{bowtie}. The circular portion 
in the figure depicts the strongly
connected component derived from $G_s$. Links out of this portion of the graph
are from people nodes and go to sinks, which are movies.

\subsection{Hammocks}
For a given recommender dataset, there are many ways of inducing the social
network graph and the recommender graph. 
The simplest, the skip, is illustrated in Figure~\ref{jc-intro}. 
Note that jumping connections provides a systematic way to
characterize recommender systems algorithms in the literature. 
We will focus on one jump called the hammock jump --- a more comprehensive
list of jumps defined by different algorithms is explored by
Mirza~\cite{batul-thesis}, we do not address them for want of space.

A hammock jump brings two people together in $G_s$
if they have at least $w$ movies in common in $\mathcal{R}$. 
Formally, a pair $(p_1, p_2)$ is in $\mathcal{J} (\mathcal{R})$ 
whenever there is a set $M_{(p_1,p_2)}$ of $w$ movies such that there
is an edge from $p_1$ and $p_2$ to each element of $M_{(p_1,p_2)}$.
The number $w$ of common artifacts is called the hammock width.
Figure~\ref{hammock-pic} illustrates a sequence (or \emph{hammock
path}) of hammocks.

There is some consensus in the community
that hammocks are fundamental in recommender
algorithms since they represent commonality of ratings.
It is our hypothesis that hammocks are fundamental to all recommender
system jumps. Early recommendation projects such as GroupLens \cite{konstan1},
LikeMinds \cite{lminds2}, and Firefly \cite{ringo1} can be viewed as 
employing (simple versions
of) hammock jumps involving at most one intermediate person.

The horting algorithm of 
Aggarwal et al.~\cite{aggarwal1} extends this idea to a sequence of hammock
jumps. Two relations --- horting and predictability ---
are used as the basis for a jump.
A person $p_1$ {\it horts} person $p_2$ if the ratings they have in common
are a sufficiently large subset of the ratings of $p_1$.
A person {\it predicts} another if they have a reverse horting relationship,
and if there is a linear transformation between their ratings.
The algorithm first finds shortest paths of hammocks that relate to
predictability and then propagates ratings using the linear
transformations.
The implementation described by Aggarwal et al.~\cite{aggarwal1} uses a 
bound on the length of the path.

There are a number of interesting algorithmic 
questions that can be studied.
First, since considering more common ratings can be beneficial (see
\cite{herlocker} for approaches) having a wider hammock could be better (this
is not exactly true, when correlations between ratings are 
considered \cite{herlocker}).
Second, many recommender systems require a minimum number $\kappa$ of
ratings before the user may use the system, to prevent {\it free-riding}
on recommendations \cite{freeriding}. What is a good value for
$\kappa$? And, third what is the hammock diameter or how far would we have
to traverse to reach everyone in the social network graph?
We begin looking at these questions in the next section.

\subsection{Random Graph Models}

Our goal is to be able to answer questions about hammock width,
minimum ratings, and path length in a typical graph.
The approach we take is to use a model of random graphs adapted from
the work of Newman, Strogatz, and Watts~\cite{newman}.
This model, while having limitations, is the best-fit of existing
models, and as we shall see, provides imprecise but descriptive 
results.

A recommender dataset $\mathcal{R}$ can be characterized by the number
of ratings that each person makes, and the number of ratings that each
artifact receives.
These values correspond to the degree distributions in the bipartite
rating graph for $\mathcal{R}$.
These counts are relatively easy to obtain from a dataset and so could
be used in analysis of appropriate algorithms.
Therefore, we would like to be able to characterize a random bipartite graph
using particular degree distributions.
This requirement means that the more common 
random graph models (e.g., ~\cite{erdos}) are
not appropriate, since they assume
that edges occur with equal probability.
On the other hand, a model recently proposed
by Aiello, Chung, and Lu~\cite{call-graph} is
based on a power-law distribution, similar to characteristics
observed of actual recommendation datasets (see next section).
But again this model is not directly parameterized by the degree
distribution. 
The Newman-Strogatz-Watts model is the only (known) model that
characterizes a family of graphs in terms of degree distributions. 

From the original bipartite graph $G = (P \cup M, E)$ for
$\mathcal{R}$ we develop two models, one for the social network graph
$G_s$ and one for the recommender graph $G_r$. 

\subsection{Modeling the Social Network Graph} Recall that the social
network graph $G_s = (P,E_s)$ is undirected and $E_s$ is induced by a jump
function $\mathcal{J}$ on $\mathcal{R}$. 
The Newman-Strogatz-Watts model works by characterizing the degree
distribution of the vertices, and then using that to compute the
probability of arriving at a node.
Together they describe a random process of following a path through
a graph, and allow computations of the length of paths.
Here we only discuss
the equations that are used, and not 
the details of their derivation.
The application of these equations to these graphs is outlined by
Mirza~\cite{batul-thesis} and is based on the derivation by Newman et
al.~\cite{newman}.

We describe the social network graph $G_s$ by the probability that a
vertex has a particular degree.
This is expressed as a generating function $G_0 (x)$ 
$$G_0 (x) = \sum_{k=0}^{\infty} p_k x^k, $$
where $p_k$ is the probability that a randomly chosen
vertex in $G_s$ has degree $k$.
This function must satisfy the property that
$$G_0 (1) = \sum_{k=0}^{\infty} p_k = 1.$$

To obtain an expression that describes the typical length of a path,
we can consider how many steps we need to go from a node to be able to
get to every other node in the graph.
To do this we can use the number of neighbors $k$ steps away.
For a randomly chosen vertex in this graph, $G_0 (x)$ gives us the
distribution of the immediate neighbors of that vertex. 
So, we can compute the average number of vertices $z_1$ one edge away from a
vertex as the average degree $z$: 
$$z_1 = z = \sum_{k} k p_k = G_0^{'} (1) $$
The number of neighbors two steps away is given by 
$$z_2 =  \sum_{k} k p_k {1 \over z} \sum_{k} k (k-1) p_k$$
It turns out (see~\cite{newman} for details) that the number of neighbors
$m$ steps away is given in terms of these two quantities:
$$z_m = {\left( {z_2 \over z_1} \right) }^{m-1} z_1$$
The path length $l_{pp}$ we are interested in is the one that is big enough to
reach all of the $N_P$ elements of $P$, and so $l_{pp}$
should satisfy the equation
$$1 + \sum_{m=1}^{l_{pp}} z_m = N_P $$
where the constant $1$ counts the initial vertex.
Using this equation, it can be shown that the typical length from one
node to another in $G_s$ is
\begin{equation}
l_{pp} = {{log[(N_P -1) (z_2 - z_1) + z_1^{2}] - log[z_1^{2}]} \over 
{log[z_2/z_1]}} \label{length1}
\end{equation}
We use this formula as our primary means of computing the
distances between pairs of people in $G_s$ in the empirical evaluation
in the next section.
Since we use actual datasets, we can compute $p_k$ as the fraction of
vertices in the graph having degree $k$. 

\subsection{Modeling the Recommender Graph} 
The recommender graph $G_r = (P \cup M,E_{sd} \cup E_{md})$ 
is directed, and hence the generating function for vertex degrees
should capture both indegrees and outdegrees: 
$$G (x,y) = \sum_{j=0,k=0}^{j=\infty,k=\infty} p_{jk} x^j y^k,$$
where $p_{jk}$ is the probability that a randomly chosen vertex has
indegree $j$ and outdegree $k$. 

From the jumping connections construction, we know that movie vertices
have outdegree $0$ (the converse is not true, vertices with outdegree $0$
could be people nodes isolated as a result of a severe jump constraint).
Notice also that by using the joint distribution $p_{ij}$, independence of
the indegree and outdegree distributions is \emph{not} implied. 
We show in the next section that this feature is very useful. 
In addition, the average number of arcs entering (or leaving) a
vertex is zero. 
And, so 
$$\sum_{jk} (j-k) p_{jk} = \sum_{jk} (k-j) p_{jk} = 0.$$
We arrive at new expressions for $z_1$ and
$z_2$~\cite{batul-thesis}: 
\begin{eqnarray*}
z_1 & = & \sum_{jk} k p_{jk}. \\
z_2 & = & \sum_{jk} j k p_{jk}. 
\end{eqnarray*} 
The average path length $l_r$
can be calculated as before: 
\begin{equation}
l_r = {{log[(N_P + N_M -1) (z_2 - z_1) + z_1^{2}] - log[z_1^{2}]} 
\over {log[z_2/z_1]}} \label{length2},
\end{equation}
where $N_P + N_M$ is the size of the recommender graph $G_r$ (assuming that 
the graph is one giant component), 
with $N_M$ denoting the number of movies.
The length $l_r$ includes paths from people to movies, as
well as paths from people to people. 
The average length of only reaching movies from people 
$l_{pm}$ can be expressed as: 
\begin{equation}
l_{pm} = {{(l_r (N_P (N_P -1) + N_P N_M) - l_{pp} N_P (N_P -1))} \over {N_P N_M}} \label{length3}
\end{equation}

\subsection{Caveats with the Newman-Strogatz-Watts Equations} There are various
problems with using the above formulas in a realistic setting
\cite{heath1}. First, unlike most results in random graph theory, the
formulas
do not include any guarantees and/or confidence levels.  Second, all
the equations above are obtained over the ensemble of random graphs that
have the given degree distribution, and hence assume that all such graphs
are equally likely.  The specificity of the jumping connections construction implies that the
$G_s$ and $G_r$ graphs are poor candidates to serve as a typical random
instance of a graph.

In addition, the equations utilizing $N_P$ and $N_M$ assume that all nodes
are reachable from any starting vertex (i.e., the graph is one giant
component). This will not be satisfied for very strict jumping constraints.
In such cases, Newman, Strogatz, and Watts suggest the substitution of these
values with measurements taken from the largest component of the graph.
Expressing the size of the components of the graph using generating
functions is also suggested \cite{newman}. However, the complexity of jumps
such as the hammock can make estimation of the cluster sizes extremely
difficult, if not impossible (in the Newman-Strogatz-Watts model). We leave this issue to
future research.

Finally, the Newman-Strogatz-Watts model is fundamentally more
complicated than traditional models of random graphs. 
It has a potentially infinite set of parameters
($p_k$), doesn't address the possibility of multiple edges, loops and, by
not fixing the size of the graph, assumes that the same degree distribution
sequence applies for all graphs, of all sizes. 
These observations hint that we cannot hope for more than a
qualitative indication of the dependence of the average path length on
the jump constraints. 
In the next section, we describe how well these formulas perform on
two real-world datasets. 

\section{Experimental Results}
\label{results}

We devote this section to an investigation of two actual datasets from the
movies domain; namely the EachMovie dataset, collected by the Digital
Equipment Corporation (DEC) Systems Research Center, and the MovieLens \cite
{movie1} dataset developed at the University of Minnesota. Both these datasets
were collected by asking people to log on to a website and rate movies.
The time spent rating movies was repaid by providing predictions of
ratings for other movies not yet seen, which the recommendation engines
calculated based on the submitted ratings and some other statistical
information. 

The datasets contain some basic demographic information about the people
(age, gender, etc) as well the movies (title, genre, release date, etc).
Associated with each person and movie are unique $id$s. The rating
information (on a predefined numeric scale) is provided as a set of
$3-$tuples: the person $id$, the movie $id$, and the rating given for that
movie by that person.  Some statistics for both the datasets are provided in
Table~\ref{stats1}. Notice that only a small number of actual ratings are
available (as a fraction of all possible combinations), and yet the
bipartite graphs of people versus movies are connected, in both cases. 

\subsection{Preliminary Investigation}
\begin{table}
\caption {Some statistics for the EachMovie and MovieLens datasets.}
\vspace{1.5mm}
\centering
\begin{tabular}{|c|c|c|c|c|} \hline\hline
\emph{Dataset} & \emph {Number of people} & \emph{Number of movies} &
\emph{Sparsity} & \emph{Connected?} \\
\hline MovieLens & 943 & 1,682 & 93.70\% & Yes\\ 
\hline EachMovie & 61,265 & 1,623 & 97.63\% & Yes\\ \hline
\end{tabular}
\label{stats1}
\end{table}

Both the EachMovie and MovieLens datasets
exhibit a {\it hits-buffs} structure. Assume that people are ordered
according to a {\it buff index} $b$: A person with buff index $1$ has seen
the most number of movies, and so on. 
For example, in the EachMovie
dataset, the person with buff index $1$ has seen $1,\! 455$ movies from 
the total of
$1,\! 623$. These $1,\! 455$ movies have, in turn, been seen by $61,\! 249$ other
people. Thus, within two steps, a total of $62,\! 705$ nodes in the graph can
be visited; with other choices of the intermediate buff node, the entire
graph can be shown to be connected in, at the most, two steps. The MovieLens
dataset satisfies a similar property. 

Furthermore, the relationship between the buff index
$b$ and the number of movies seen by the buff $P(b)$
follows a power-law distribution, with an exponential cutoff:
$$P(b) \propto {b}^{- \alpha} e^{-{{b}\over{\tau}}}$$ 
For the EachMovie dataset, $\alpha \approx 1.3$ and $\tau \approx 10,\! 000$.
Similar trends can be observed for the hits and for the MovieLens
dataset. Graphs with such power-law
regimes can thus form small-worlds \cite{small-world-classes} as
evidenced by the short length between any two people in both MovieLens and
EachMovie. 
To better demonstrate
the structure, we reorder the people and movie ids, so that the
relative positioning of the ids denotes the extent of a person being a
buff, or a movie being a hit. For example, person id $1$ refers to the
person with buff index $1$ and movie id $1$ refers to the movie with hit
index $1$.
Figure~\ref{movielens-matview} illustrates the hits-buffs structure of 
the MovieLens dataset.

\begin{figure}
\centering
\begin{tabular}{cc}
\end{tabular}
\caption{Hits-buffs structure of the (reordered) MovieLens dataset.}
\label{movielens-matview}
\end{figure}

\begin{figure}
\centering
\begin{tabular}{cc}
& \mbox{\psfig{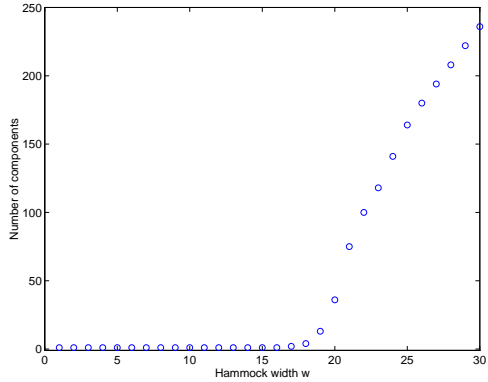}}
\end{tabular}
\caption{Effect of the hammock width on the number of components in the 
$G_r$ graph induced from the MovieLens dataset.}
\label{movielens1}
\end{figure}

\subsection{Experiments}
The goal of our experiments is 
to investigate the
effect of the hammock width $w$ on the average characteristic path lengths
of the induced $G_s$ social network graph
and $G_r$ recommender graph for the above datasets. 
We use versions of the EachMovie and MovieLens datasets sanitized by removing
the rating information. So, even though rating information can easily be used by
an appropriate jump function (an example is given in~\cite{aggarwal1}), we
explore a purely connection-oriented jump in this study.
Then, for various values of the hammock
width $w$, we form the social network and recommender graphs and
calculate the degree
distributions (for the largest connected
component). This was used to obtain the lengths predicted by equations
~\ref{length1} and ~\ref{length2}
from Section~\ref{jcmodel}. We also compute
the average
path length for the largest connected component of both the secondary graphs 
using parallel implementations of 
Djikstra's and Floyd's algorithms. The experimental observations are compared
with the formula predictions.

\subsubsection{MovieLens} 
\begin{figure}
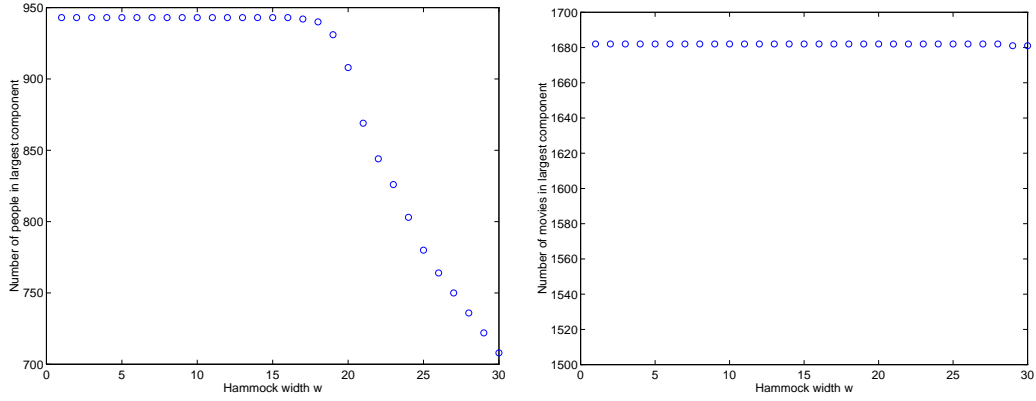

\centering
\begin{tabular}{cc}
\mbox{\psfig{figure=movielens2.epsi,width=2.6in}} &
\mbox{\psfig{figure=movielens3.epsi,width=2.6in}}\\
\end{tabular}
\caption{(left) Effect of the hammock width on the number of people in the
largest components in the MovieLens $G_r$ graph. (right)
Effect of the hammock width on the number of movies in the
largest components in the MovieLens $G_r$ graph.}
\label{movielens2}
\end{figure}

Fig.~\ref{movielens1} 
describes the number of
connected components in $G_r$ as a result of imposing increasingly strict hammock jump
constraints. Up to about $w=17$, the graph remains in one piece and rapidly
disintegrates after this threshold. The value of this transition
threshold 
is not surprising, since the designers of MovieLens insisted that
every participant rate at least $\kappa = 20$ movies. As observed from
our experiment results, after the threshold 
and up to $w=28$, there is still only one giant component with isolated
people nodes (Fig.~\ref{movielens2}, left). Specifically, the degree 
distributions of the
MovieLens social network graphs for $w > 17$ show us that the 
people nodes that
are not part of the giant component do not form any other connected components
and are isolated. We say that a jump {\it shatters} a set of nodes if
the vertices that are
not part of the giant component do not have any edges.
This aspect of the formation of a giant component is well known 
from random graph
theory \cite{bollabas}. Since our construction views the movies as a
secondary mode, we can ensure that only the strictest hammock jumps
shatter
the $N_M$ movie nodes. Fig.~\ref{movielens2} (right) demonstrates that the movie
nodes are not stranded as a result of hammock constraints up to hammock
width $w=29$.


\begin{figure}
\centering
\begin{tabular}{cc}
\mbox{\psfig{figure=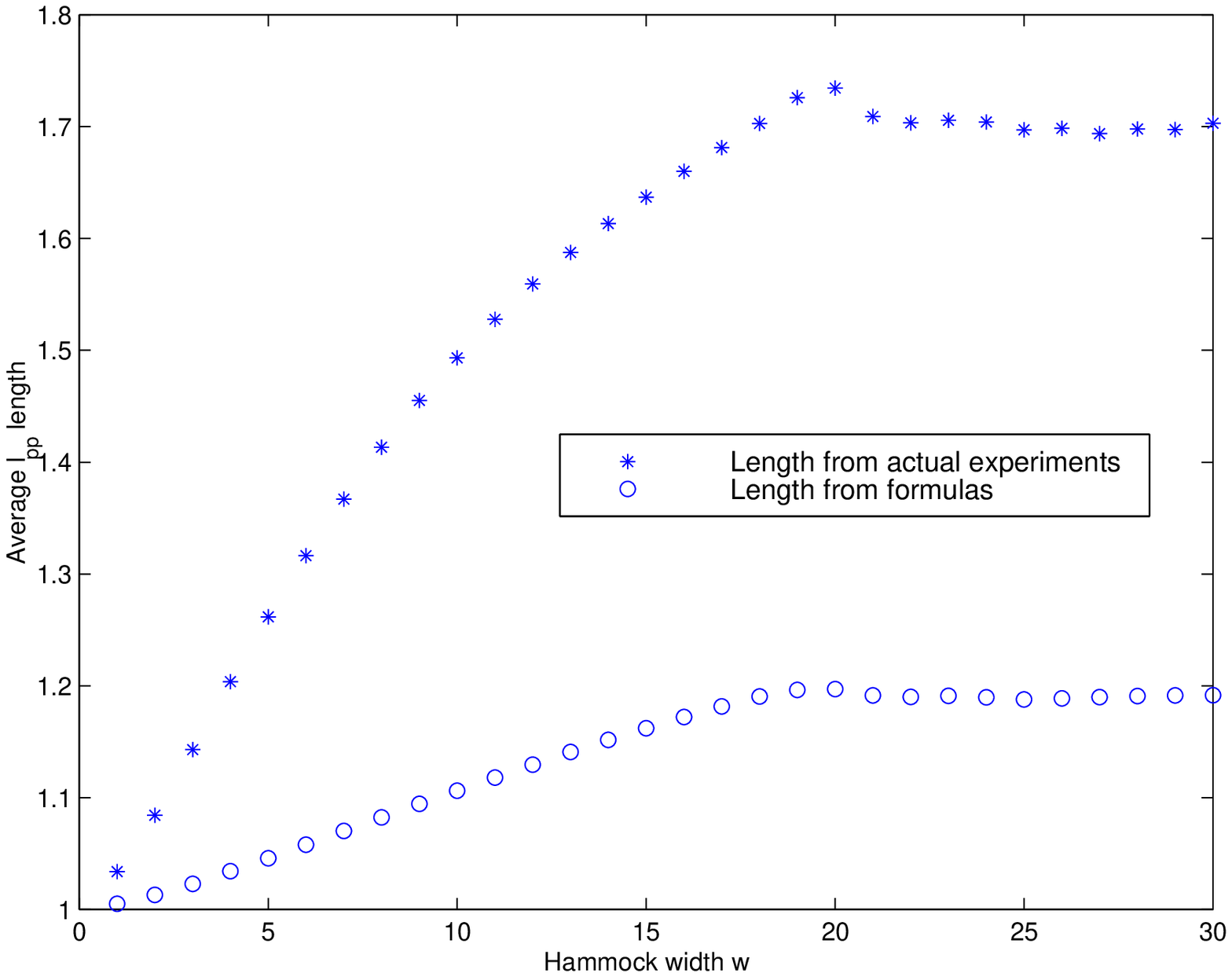,width=2.6in}} &
\mbox{\psfig{figure=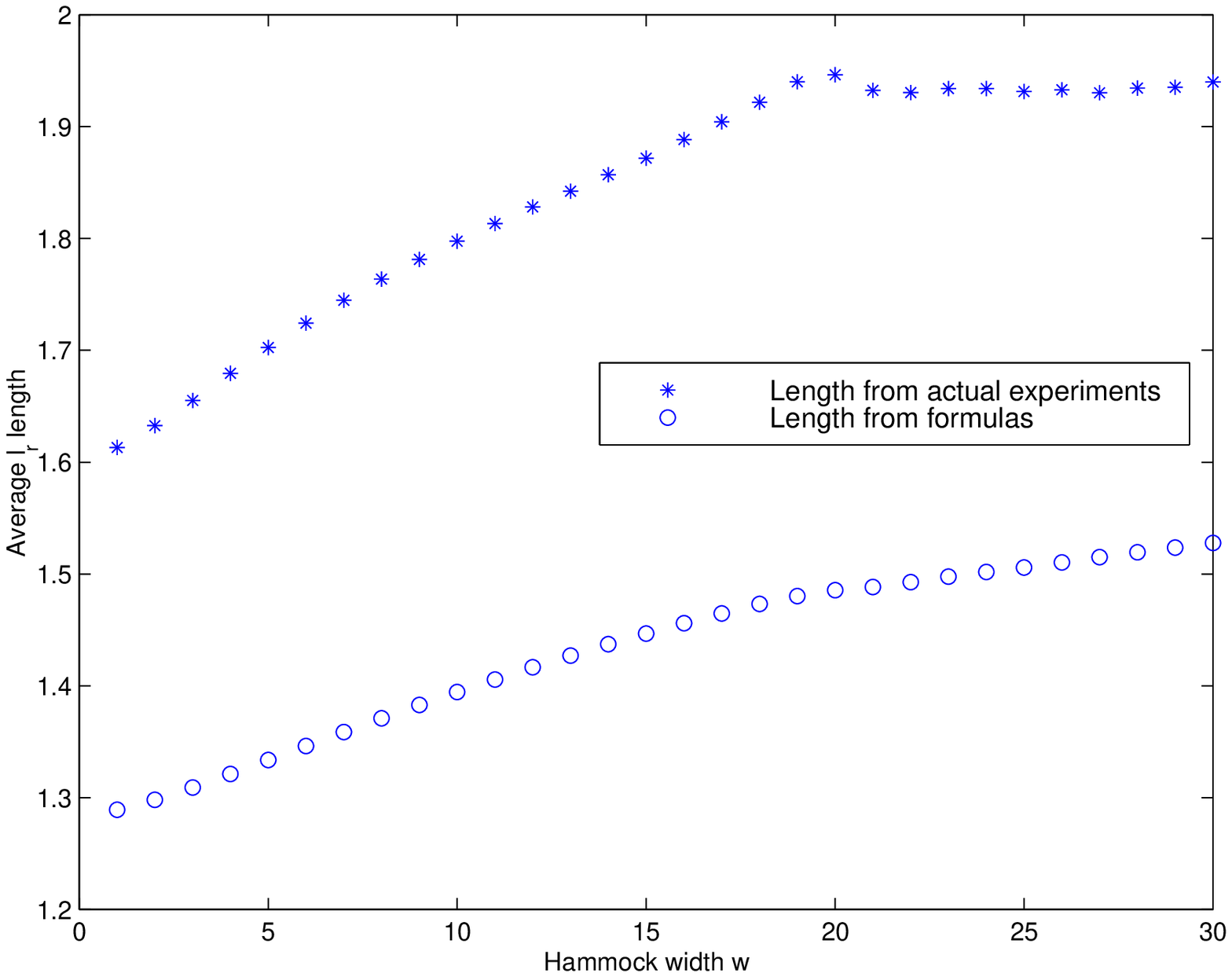,width=2.6in}}\\
\end{tabular}
\caption{(left) Comparison of the $l_{pp}$ measure (MovieLens) from actual
computations and from the formulas. (right)
Comparison of the $l_{r}$ measure (MovieLens) from actual
computations and from the formulas.}
\label{movielens4}
\end{figure}

The comparison of experimental observations with formula predictions
for the $l_{pp}$ and $l_r$ lengths are shown in
Fig.~\ref{movielens4}. The graphs for $l_{pp}$ share certain important
characteristics.
The increase in length up to the threshold point is explained by the fact
that edges are removed 
from the giant component, meaning paths of greater length have to be
traversed to reach other nodes. After the threshold, the relative
stability of the length indicates that the only edges lost are those
that are associated with the stranded people nodes.  We attribute the
fact that the lengths are between $1$ and $2$ to the hits-buffs
structure, which allows short paths to almost any movie.
Notice that while the formulas capture the qualitative
behavior of the effect of the hammock width, it is obvious from
Fig.~\ref{movielens4} (left) that they postulate significantly less clustering 
than is actually observed. A possible explanation is given later in this 
section.

The comparison of the experimental observations and formula
predictions for $l_r$ (Fig.~\ref{movielens4}, right) show
substantially
better agreement, as well as tracking of the
qualitative change. Once again, the formulas assume significantly less
clustering than the actual data. In other words, the higher values of
lengths from the actual measurements indicate that there is some source of
clustering that is not captured by the degree distribution,
and so is not included in the formulas.


\subsubsection{EachMovie}
\begin{figure}
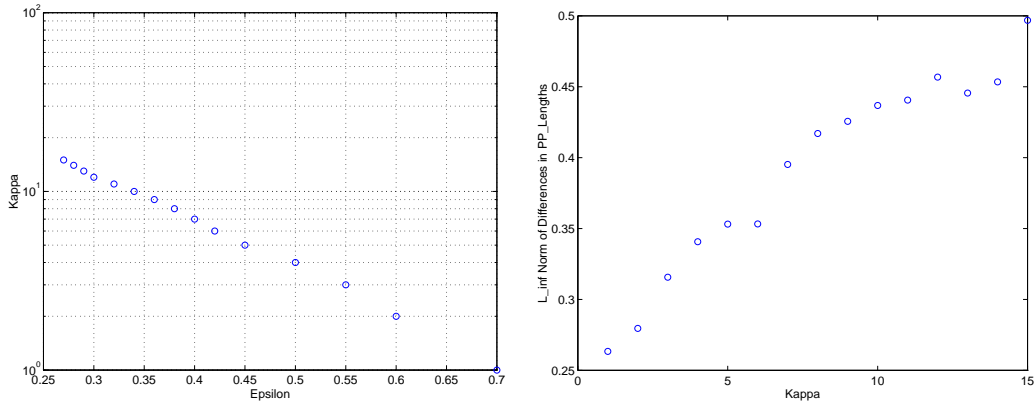

\centering
\begin{tabular}{cc}
\mbox{\psfig{figure=calib.epsi,width=2.6in}} &
\mbox{\psfig{figure=linfpp.epsi,width=2.6in}}\\
\end{tabular}
\caption {(left) Calibrating the value of $\epsilon$ in the synthetic model of
EachMovie produces datasets with
required specifications on minimum rating $\kappa$.
(right) Comparison of the $l_{pp}$ measure (EachMovie) from actual
computations and from the formulas for varying values of $\kappa$.}
\label{calibs}
\end{figure}

The evaluation of the EachMovie data is more difficult owing to
inconsistencies
in data collection. For example, the dataset was collected in two
phases, with entirely different rating instructions and scales in the
two situations, and also contains duplicate ratings. We 
concentrate on the portion of the data collected in $1997$ and
use a synthetic dataset that has the same sparsity and
exponent of the power-law as this
reduced dataset. Specifically, our dataset includes
500 people and 75 movies and has
the property that the person with buff index $b$ has seen the 
first $\lceil 75
b^{-\epsilon} \rceil$ movies (recall that the movies are also ordered according to
their hit $id$). An $\epsilon = 0.7$ produces a dataset with a minimum
rating
of $1$ movie ($95.5\%$ sparse), while $\epsilon = 0.27$ produces a minimum 
rating of $15$ movies (with sparsity $76.63\%$). The choice of $\epsilon$
thus
provides a systematic way to analyze the effect of the minimum rating
constraint $\kappa$ (see Fig.~\ref{calibs}, left). In addition, for each 
(person, movie) edge
of these synthetic graphs, we generate a uniform (discrete)
random variate in $[0,10]$ and rewire the movie endpoint of the edge 
if this variate is $< 2$. This device models deviations from a strict
hits-buffs distribution. We generated $15$ such graphs and ensured that they
were connected (in some cases, manual
changes were made to ensure 
that the graph was connected). These graphs
served as the starting points for our analysis. 
For each of these $15$ graphs, we vary the hammock width
$w$ from $1$ to $25$ and repeat the length calculations (using both
experiments and formula predictions) for the social network and
recommender graphs.

Like the MovieLens experiment, the formulas for EachMovie predict shorter
$l_{pp}$ lengths (and consequently, lesser clustering) than observed from
actual experiments. To characterize the mismatch, for each of the $15$ values
of $\kappa$, we express the differences between the formula predictions
and experimental observations as a vector (of length $25$, for each of the
$25$ values of hammock width $w$). The $L_{\infty}$ norm of this vector
is plotted against $\kappa$ in 
Fig.~\ref{calibs} (right).
Notice the relatively linear growth of the discrepancy
as $\kappa$ increases. 
In the range of $\kappa$ considered, the hammock
constraints shatter the graph into many small components, so the formulas
are applied over ever-decreasing
values of $n$, rendering them ineffective. For example,
for $\kappa = 15$, a hammock width $w = 25$ shatters the graph into
$53$ components. Since the hammock width $w$ varies in 
$[1,25]$ over a range of $[1,15]$ for $\kappa$, we have reason to believe 
this graph will taper off when extrapolated to higher values of $\kappa$. While
this experiment does not provide any new insight, it hints at a fairly
bounded growth in the $L_{\infty}$ discrepancies for $l_{pp}$ lengths.

\begin{figure}
\centering
\begin{tabular}{cc}
\mbox{\psfig{figure=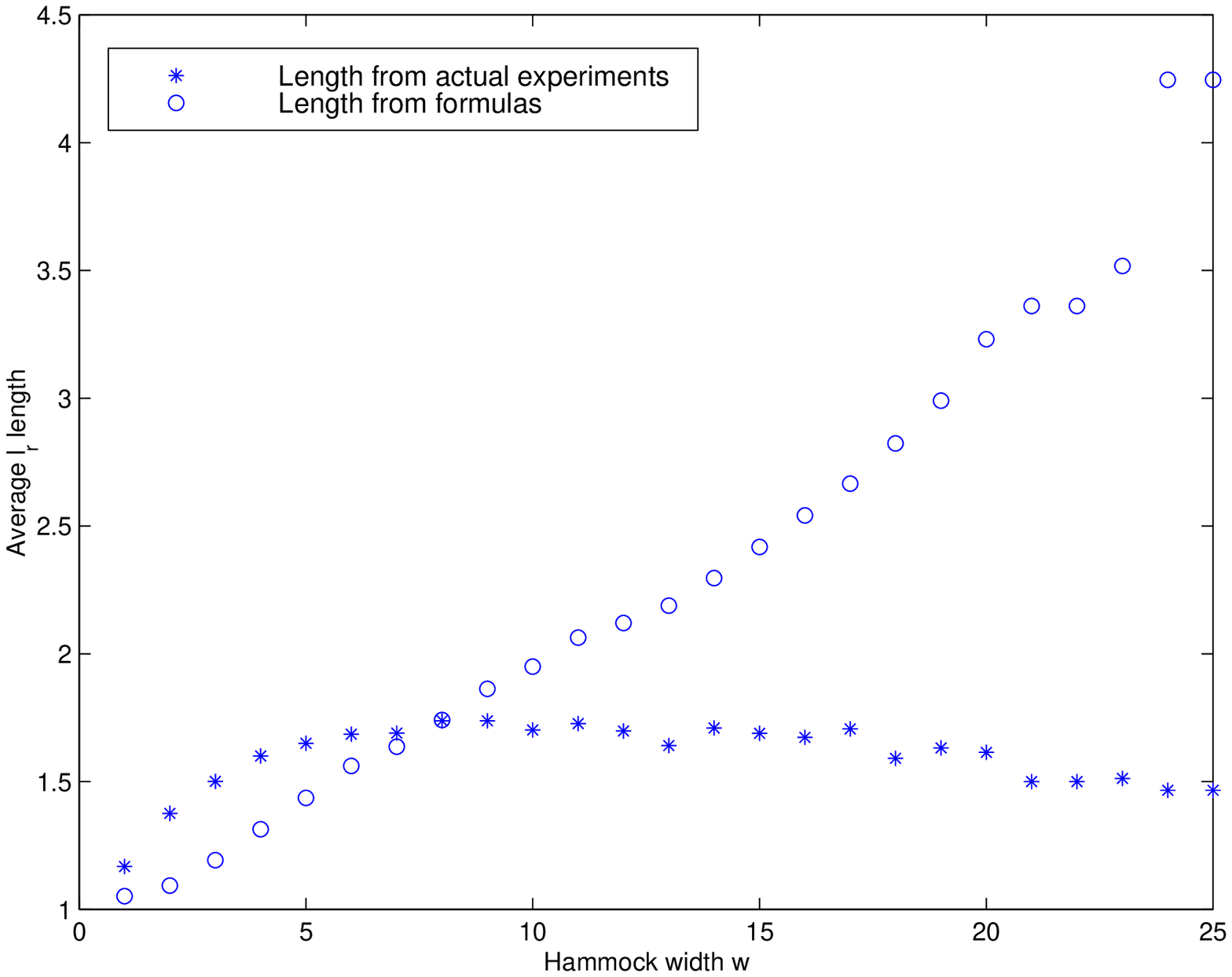,width=2.3in}} &
\mbox{\psfig{figure=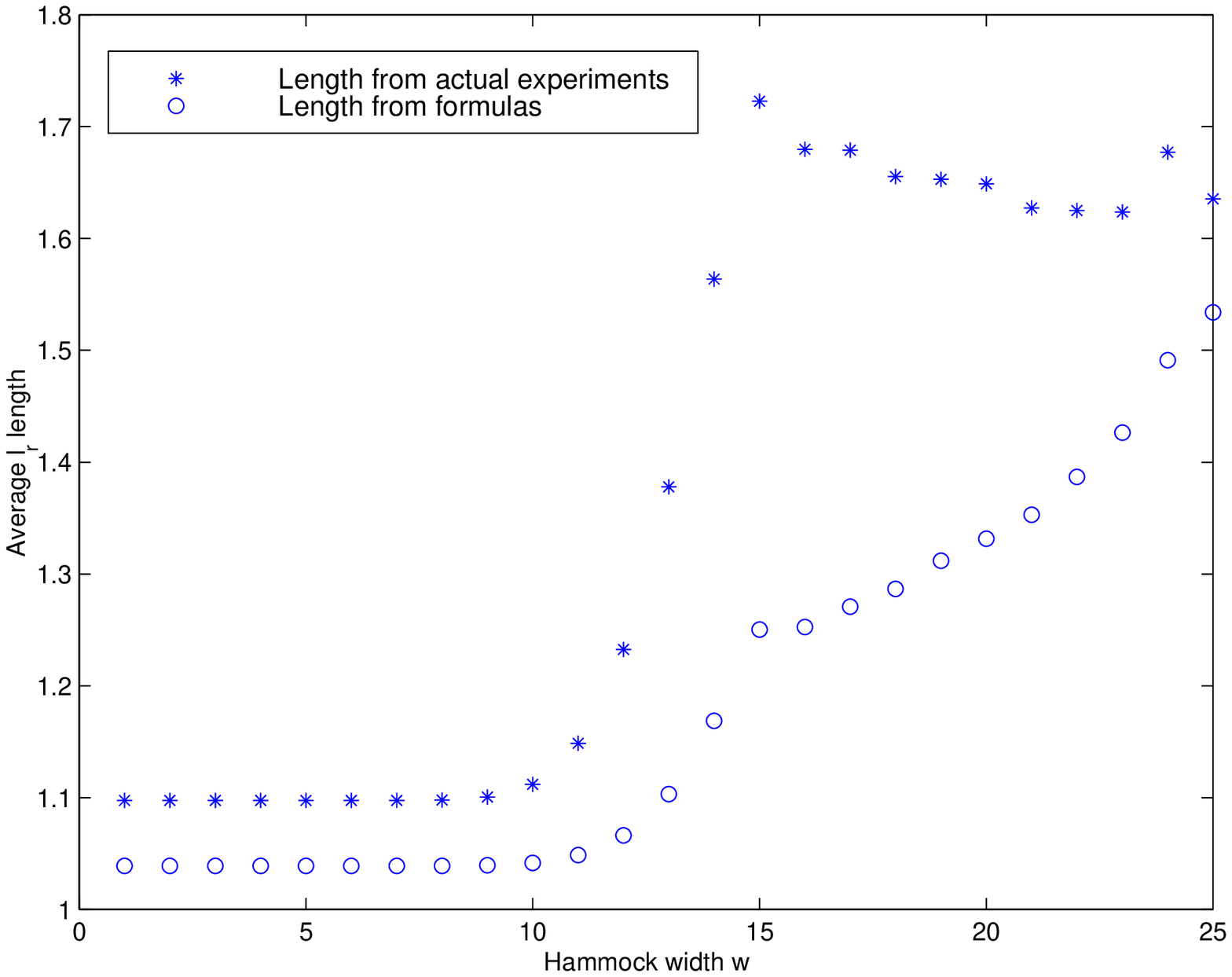,width=2.3in}}\\
\end{tabular}
\caption{Comparison of the $l_{r}$ measure (EachMovie) from actual 
computations and from the formulas for 
minimum rating constraint $\kappa = 1$ (left) and minimum
rating constraint $\kappa = 15$ (right).}
\label{lpmtrouble}
\end{figure}

The comparisons for the $l_{r}$ lengths tell a different story
(see Fig.~\ref{lpmtrouble}). At low values of $\kappa$, the $l_{r}$
length calculated from formulas is highly erroneous for even
medium values of hammock width $w$ (Fig.~\ref{lpmtrouble},
left) whereas we see the now familiar
effect of assuming too little
clustering 
for high values ($\geq 12$) of $\kappa$ (Fig.~\ref{lpmtrouble}, right). In
particular, for
$\kappa = 1, w=25$, the formulas predict an average $l_{r}$ length
of $4.24$! This is counter-intuitive given that the largest component
for this shattering itself consists of only $4$ people nodes (and $71$
movie nodes). 

\begin{figure}
\centering
\begin{tabular}{cc}
& \mbox{\psfig{figure=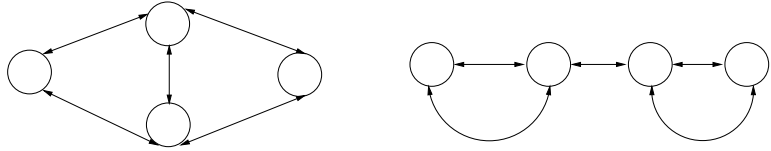,width=4in}} \\
\end{tabular}
\caption{Two graphs that satisfy the degree distribution given in
Table.~\ref{degsurprise}. For simplicity, only the people nodes are shown.}
\label{figsurprise}
\end{figure}

This problem arises because
the Newman-Strogatz-Watts
model does not prohibit a multiplicity of edges between a pair
of nodes. Let us look
at this situation more closely. For $\kappa = 1, w=25$, the largest
component has the degree distribution given in Table.~\ref{degsurprise}.
Notice that the nodes with outdegree $0$ are obviously the movie nodes
and the nodes with non-zero outdegree are the people nodes. Thus, two of the 
people are each connected to two people, and two of the people are each
connected to
three people. A graph that satisfies this property is shown
in the left of Fig.~\ref{figsurprise}. The origin of the value of $4.24$ can
be traced back, rather to the Newman-Strogatz-Watts
model's postulation of the unlikely 
culprit graph shown
in the right of Fig.~\ref{figsurprise}. Notice that this graph satisfies
the exact same distribution, but by allowing multiple edges many movie
nodes are counted more than once, and with greater (ever increasing)
path lengths. One cycle between two people can thus effect
two extra hops in the length calculations, rendering the estimates
inaccurate. 
As observed from our results, as $\kappa$ increases, the largest component
increases in size. For example, when $\kappa = 15$ and $w = 25$, the largest
component has 53 people whereas for $\kappa = 1$ and $w = 25$, there are only
4 people in the largest component. With increase in largest
component size for higher values of $\kappa$, the
proportion of such
pathological
graphs decreases; hence
the observed (qualitative) agreement between actual and predicted values.

\begin{table}
\caption {Joint degree distribution for the largest component of the recommender
graph (EachMovie) when $\kappa = 1, w=25$. Only the non-zero entries are shown.}
\vspace{5mm}
\centering
\begin{tabular}{|c|c|c|} \hline\hline
indegree $j$ & outdegree $k$ & $p_{jk}$\\ \hline
 1 & 0 & 23/75 \\
 2 & 0 & 16/75  \\
 3 & 0 & 13/75 \\
 4 & 0 & 19/75 \\ \hline
 2 & 31 & 1/75 \\
 2 & 65 & 1/75 \\
 3 & 37 & 1/75 \\
 3 & 47 & 1/75 \\ \hline
\end{tabular}
\label{degsurprise}
\end{table}

\subsection{Discussion of Results}
We can make several preliminary observations from the results so far:
\begin{enumerate}
\item As Newman, Strogatz, and Watts point out \cite{newman}, the random
graph model defined by degree distributions makes strong qualitative
predictions of actual lengths, using only local information about
the number of first and second nearest neighbors ($z_1$ and $z_2$
from Section~\ref{jcmodel}). We have demonstrated
that this holds true even for graphs
induced
by hammock jumps.
\item For sufficiently large values of $\kappa$,
the relationship between hammock width $w$ and average $l_{pp}$
length follows two distinct phases: (i) in the first regime, there
is a steady increase of $l_{pp}$ up to a threshold $< \kappa$, where
only edges are lost; (ii) in the second phase, nodes are shattered but
without much effect on the average $l_{pp}$ values. This two-phase
phenomenon can thus serve as a crucial calibration mechanism for connecting
the number of people to be brought together by a recommender system
and the average $l_{pp}$
length. For example, one can define a parametric study such as discussed
here for a new domain and proceed to first demarcate the phases. Choices of
hammock width $w$ can then be made, depending on the feasibility of
realizing an appropriate $\kappa$ (in real-life) and any desired
constraints on $l_{pp}$. 

\item Visualizing the overlap in the social network graphs,
as hammock width constraints are increased, leads to interesting observations.
As can be seen in the Venn diagram of Fig.~\ref{looping} (left), the 
composition of
the nodes in $G_s$ appears to be fairly robust. Smaller and smaller graphs
are created, but with a common core, as illustrated
by the `clam shells' in the diagram. In other words, there is not 
a handful of people or movies who form a cutset for the graph --- 
there is no small group of people or movies
whose removal (by a sufficiently strict hammock width)
causes the entire graph to be broken into two
or more pieces (see right of Fig.~\ref{looping}).  This confirms observations made elsewhere \cite{barba-albert} 
that power-laws are a major factor ensuring the robustness and scaling 
properties of graph-based networks. Our study is the first to investigate
this property for graphs induced by hammock jumps. However, we believe this
robustness is due to the homogeneous nature of ratings in the movie domain.
In other domains such as music CDs, where people can be partitioned by
preference, the diagram on the right of Fig.~\ref{looping} would
be more common.

\begin{figure}
\centering
\begin{tabular}{cc}
& \mbox{\psfig{figure=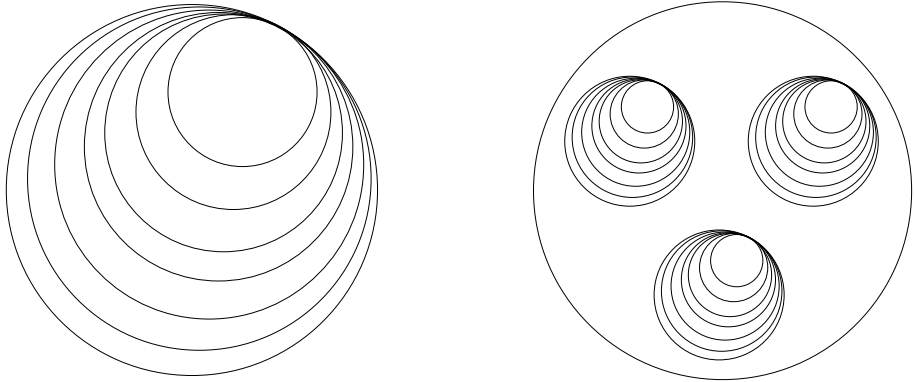,height=2in}}
\end{tabular}
\caption{(left) `Clam shells' view of the social networks $G_s$ induced by 
various 
hammock widths. Increasing hammock width leads to smaller networks. (right)
a hypothetical situation likely to be caused by the presence of a few strong
points in the graph.}
\label{looping}
\end{figure}

\item The average $l_{r}$ lengths are within the range
$[1,2]$ in both formula predictions and experimental results.
Caution has to be exercised whenever the
graph size $n$ gets small, as Fig.~\ref{figsurprise} shows. In our case,
this happens when the hammock width constraint $w$ rises 
above the minimum rating constraint $\kappa$. Of course,
this behavior should also be observed for other strict jumps.
\item Both of the $l_{pp}$ and $l_{r}$ formulas postulate consistently less
clustering than observed in the real data.  We attempt to address this
below.
The typical random graph model has a Poisson
distribution of edges \cite{bollabas}, whereas, as seen earlier,
real datasets exhibit a power-law 
distribution \cite{doyle-ref,falout}. The power-law feature is sometimes
described as `scale-free' or `scale-invariant' since a single parameter 
(the exponent of the law) captures the size of the system at all stages in
its cycle (the x-axis of the law). A log-log plot of values would thus
produce a straight line, an effect not achievable by traditional random
graph models. Barab\'{a}si and Albert \cite{barba-albert} provide two
sufficient conditions for this property: growth and
preferential attachment. {\it Growth} refers to the ability of the system
to dynamically add nodes; random graph models that fix the number
of nodes are unable to expand. {\it Preferential attachment} refers to
the phenomenon that nodes that have high degrees have a greater propensity
of being linked to by new nodes. In our case, a movie that is adjudged
well by most people is likely to become a hit when additional people are
introduced.  Barab\'{a}si and Albert
refer to this as a ``rich get richer'' effect \cite{barba-albert}.

\begin{figure}
\centering
\begin{tabular}{cc}
\end{tabular}
\caption{Cumulative frequency distribution (of degrees) in $G_s$
as a function of
the degree for (left) MovieLens and (right) EachMovie datasets, for 
hammock widths from $1$ to $30$.} 
\label{cumfreq1}
\end{figure}

\begin{figure}
\centering
\begin{tabular}{cc}
\end{tabular}
\caption{Logarithm of the cumulative frequency distribution (of degrees) in
$G_s$ as a function of
the degree for (left) MovieLens and (right) EachMovie datasets, for 
hammock widths from $1$ to $30$.}
\label{cumfreq2}
\end{figure}

To characterize the possible causes in our domain, we consider the distribution
of degrees in the social network graphs $G_s$ of both MovieLens and
(the actual) EachMovie datasets (see Fig.~\ref{cumfreq1}). In the figure, 
lines from top to bottom indicate increasing
hammock widths. 
Both datasets do not follow a strict power law
for the entire range of the hammock width $w$.
For low values of $w$, there is a small and steadily
increasing power-law regime, followed by an exponential cutoff. For higher
values of $w$, the graph `falls' progressively earlier and the CDF
resembles a Gaussian or exponential decay, with no power-law behavior.
This is evident if the logarithm of the CDF is plotted, as in
Fig.~\ref{cumfreq2} where top to bottom indicates increasing
hammock widths. Notice the
significant cusp in the left side of both
graphs, depicting the qualitative change in behavior.
These two extremes of
deviations from power-law behavior (at low and high values of $w$)
are referred to as
{\it broad-scale} and {\it single-scale} \cite{small-world-classes},
respectively,
to distinguish them from the scale-free behavior of power-laws. 

Causes for such deviations from a pure power-law are also understood,
e.g. aging and capacity.
{\it Aging} refers
to the fact that after a certain point in time, nodes stop accumulating
edges. {\it Capacity} refers to resource-bounded environments where cost and
economics prevent the hits from becoming arbitrarily greater hits. These
observations have been verified
for environments such as airports (capacity
limitations
on runways), and natural networks (aging caused by people dying) 
\cite{small-world-classes}.

In our domain, recall that the $G_s$ graphs model the connectivities among
people, and are hence indirect observations from
an underlying bipartite graph.  One possible explanation for the deviations
of the connectivities in $G_s$ from
from power-law behavior is suggested in a study done by Robalino and Gibney
\cite{Robalino}, which models the impact of movie demand on the social
network underlying a word-of-mouth recommendation. The two new factors
introduced there are {\it expectation} and {\it homogeneity} (of the social
network). The authors suggest that ``some movies might end up having low
demand, depending on the initial agents expectations and their propagation
through the social network'' \cite{Robalino}. 
Furthermore, they assume that 
negative
information (ratings) obtained early in a movie's lifetime can have
substantial effect in a homogeneous network (one where individuals
trust each others opinions strongly than in other networks). At this point,
we are unable
to accept or reject this observation due to lack of information about the
social dynamics in which the data was collected in MovieLens and EachMovie.

However, such an effect might not still model
the shift in the emphasis from a broad-scale
behavior to a single-scale behavior as $w$ increases. Fortunately, this
is easy to explain algorithmically from
our construction. For higher values of $w$, the degree
distributions resemble more and more a typical random graph (for smaller
values of $n$), which has
a connectivity characterized by a fast decaying tail (such as a Poisson
distribution). 
Insisting on greater
values of $w$ leads to higher and higher decays, such that for sufficiently
large $w$ (relative to $\kappa$), 
no power-law regime is visible \cite{small-world-classes}.

\end{enumerate}

\section{Concluding Remarks}
\label{future}

This research makes two key contributions.
First, we have
shown how algorithms for recommender systems can be characterized as jumping
connections in a bipartite graph. This view enables
a new methodology to conduct experimental analysis and comparison of
algorithms. Using this approach, algorithms can be distinguished by the
pairs of nodes that are brought together.

Second, we have described the application of our framework
to a particular form of jump --- the hammock jump. We have demonstrated
a two-phase phenomenon for the induced social network graph that allows us
to connect the minimum rating constraint $\kappa$, the hammock width
$w$, the size of the largest component, and the average person-person
length $l_{pp}$.
In particular, the choice of $\kappa$ determines the phase transition in
varying the hammock width $w$, with $\kappa$ being an upper bound on
$w$. Once formalized further (see below), this connection will permit
tradeoff analysis between choices for the minimum rating constraint, and
the strength of jumps as measured by the hammock width.

The eventual success of the proposed methodology relies on the
expressiveness of the representations supplied to the recommender system
builder and his/her ability to reason effectively with such representations.
Ideally,  the designer
will fix one or more of the variables among $\kappa$, $w$, average
$l_{pp}$ length, and the types and numbers of nodes brought together by
a jump.
An analysis for a particular degree distribution will help make
estimates for other parameters.

Recall
the situations described in Section~\ref{intro}:
\begin{itemize}
\item {\it Scenario 1} involves finding clusters of people with related
interests. This can be done by
calibrating the hammock width using a plot
of the type shown in Fig.~\ref{movielens2} (left).
\item {\it Scenario 2} involves exploring the connectivity properties, perhaps
via visualization, of the social network graphs induced by the jumps
under consideration (see Fig.~\ref{looping}). 
\item {\it Scenario 3} involves a calibration of $\kappa$ based
on synthetic (or random) graphs generated by the expected rating patterns.
Current work is addressing random graph models that would be applicable
in this context --- see discussion of future work below.
\end{itemize}

\noindent
Note that our work emphasizes connections, or whether a recommendation is
possible. This is complementary to considering predictive accuracy, which
must be assessed by a field study. 

We now describe various opportunities for future research involving 
graph analysis, more expressive forms of jumps, and development of new
random graph models.

\subsubsection*{Role of Hits-Buffs}
The existence of ratings structures such as the hits-buffs
distribution and our ability to exploit them (to minimize factors such as
the average length) is very crucial to the success of recommender systems.
An interesting question is how much of the hits-buffs structure
should be present in a dataset to provide high quality recommendations?
The answer to this question has implications for current
methodologies of data acquisition. Typically, movies (or the
secondary mode) are
partitioned into two sets: a {\it hot set}, that almost everybody is
required to rate (to increase commonality of ratings)
\cite{aggarwal1,eigentaste} and a {\it cold set} that is used to
ensure adequate coverage \cite{aggarwal1}. A more detailed understanding of
the hits-buffs structure would provide new methodologies to address this
dichotomy.

\begin{figure} 
\centering 
\begin{tabular}{cc}
\mbox{\psfig{figure=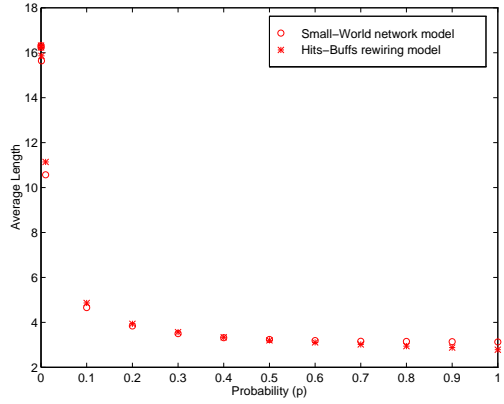,width=2.6in}} 
\end{tabular}
\caption{(left) Effect of two rewiring models on the characteristic path length,
starting from a regular wreath network.
(right) A stratified view of the MovieLens dataset demonstrates hits-buffs
structures at all rating levels.} 
\label{new-ws} 
\end{figure}

To study this question, we conducted a Watts-Strogatz analysis with a
rewiring model that preferentially rewires to some nodes with greater
probability. Thus, nodes that are hits or buffs have a greater propensity of
being linked to. Fig.~\ref{new-ws} (left) shows the results for the average 
length
in a one-mode graph. For small values
of rewiring probability $p$, the results from a
preferential rewiring are virtually indistinguishable from a random
rewiring, so both of them serve to bring the length down to nearly identical
limits. For large values of
$p$, the influence of the hits-buffs structure
is evident in the reduced length. It is thus unclear what a base
structure should be to explore the role of hits-buffs in length reduction.
Recent research on the modeling of dynamic systems shows that power-laws
such as hits-buffs are crucial to ensure robustness \cite{doyle-ref}; more
work needs to be done to delineate connections to data collection and
desired metrics in recommender systems.

\subsubsection*{Expressive Jumps}
Our experimentation has concentrated on the hammock jump varied by the
hammock width parameter.
However, it is also possible to consider other jumps within this
framework, and compare them by the connections they make for the same
recommender dataset.
This comparison can concentrate on the clusters of people (or people and
movies) that are brought together (illustrated in Fig.~\ref{looping}).
For instance, given two algorithms $\mathcal{A}$ and $\mathcal{B}$ parameterized
by $\alpha$, we would be able to make statements of the form
\begin{descit}
Algorithm $\mathcal{A}$ with $\alpha=2$ makes the same connections as
algorithm $\mathcal{B}$ with $\alpha\leq 6$.
\end{descit}
Results of this form would aid in the selection of recommendation
algorithms by the relative strengths 
of the underlying jump.

This approach would clearly point us in the direction of trying to
find algorithms that use the strongest jump possible for all
datasets.  However, these algorithms may be computationally intensive (for
instance, recently proposed
algorithms that use mixture models~\cite{Heck-New}
and latent variables~\cite{Hofmann}), while in a particular dataset the
jump is actually equivalent to one performed by a much less expensive
algorithm.
This point is illustrated by the observation in our experimentation
that the path lengths in the recommender graph are between 1 and 2,
which suggests that algorithms that attempt to find longer paths will
end up being equivalent to algorithms that do not.
Therefore, the choice of algorithm is strongly dependent on the
dataset, and simpler machinery may be more efficient and cost
effective than more complex algorithms that 
would perform better
in other contexts.

The need to identify efficient forms of jumping will become more important
when additional information, such as rating values,
is included in jump calculations. Fig.~\ref{new-ws} (right) shows the ratings 
information present in the MovieLens dataset. As can be seen, the hits-buffs 
structure gets stratified into multiple layers and can be used advantageously 
both within a ratings layer and across layers. Algorithms can exploit
such prior knowledge of rating patterns to make cheaper jumps.

Finally, we make the observation that almost always,
\emph{recommendation algorithms never bring disconnected portions of a graph 
together}, and when they do, it is in only very pathological cases.
This observation, in part, leads us to the belief that all recommender
jumps depend in some way on hammock jumps.
In some cases this is obvious, but others will require more study.

\subsubsection*{New Random Graph Models}
The formulas presented in this
paper for
$l_{pp}$ and $l_{r}$ are derived from the parameters of the induced $G_s$
and $G_r$ graphs. One possible direction of work is to cast these variables
in terms of parameters of the original bipartite graph (or dataset
$\mathcal{R}$).  However, the Newman-Strogatz-Watts
model is very difficult to analyze for all
but the simplest forms of jumps. Recall the Aiello-Chung-Lu model 
\cite{call-graph} for massive graphs modeled
after power-laws. Unlike the Newman-Strogatz-Watts
model and like traditional random graph
models, this model has only two parameters (the intercept and slope of the
power-law plotted on a log-log scale). Estimations of graph properties such
as diameter have very recently been initiated \cite{Lu--diameter} for this
new model and it appears to be a promising candidate for application to
recommender systems. In particular, we could aim for a more accurate modeling
of the connection between $\kappa$ and the hammock width constraint $w$ at
which the graph becomes shattered.\\

\noindent
{\it Acknowledgements:} The authors express their thanks to L.S. Heath
for helpful discussions. Acknowledgements are also due to the Compaq
Equipment Corporation, which provided the EachMovie dataset and the
University of Minnesota, which provided the MovieLens dataset used in our
experiments.

\bibliographystyle{plain}
\bibliography{paper}

\end{document}